\documentclass[10pt]{article}

\usepackage{amsmath}
\usepackage{amssymb}

\usepackage{graphicx}

\usepackage{cite}

\usepackage{color} 

\usepackage{url} 
\usepackage{xr} 
\usepackage[labelfont=bf,labelsep=period,justification=raggedright]{caption}

\makeatletter
\renewcommand{\@biblabel}[1]{\quad#1.}
\makeatother

\date{}

\begin{document}

\begin{flushleft}
{\Large
\textbf{Automatic Network Fingerprinting Through Single-Node Motifs}
}
\\
\vspace{3mm}
Christoph Echtermeyer$^{1}$,
Luciano da Fontoura Costa$^{2}$,
Francisco A. Rodrigues$^{3}$,
Marcus Kaiser$^{1,4,5,\ast}$
\\
\vspace{3mm}
\bf{1} School of Computing Science, Claremont Tower, Newcastle University, Newcastle-upon-Tyne NE1 7RU, UK
\\
\bf{2} Instituto de F\'isica de S\~ao Carlos,
Universidade de S\~{a}o Paulo, S\~{a}o Carlos, PO Box 369, 13560-970 S\~{a}o Carlos, S\~{a}o Paulo, Brazil
\\
\bf{3} Departamento de Matem\'{a}tica Aplicada e Estat\'{i}stica, Instituto de Ci\^{e}ncias Matem\'{a}ticas e de Computa\c{c}\~{a}o, Universidade de S\~{a}o Paulo, S\~{a}o Carlos, PO Box 668, 13560-970 S\~{a}o Carlos, S\~{a}o Paulo, Brazil
\\
\bf{4} Institute of Neuroscience, The Medical School, Framlington Place, Newcastle University, Newcastle-upon-Tyne NE2 4HH, UK
\\
\bf{5} Department of Brain and Cognitive Sciences, Seoul National University, Seoul 151-746, Korea
\\
\vspace{3mm} 
$\ast$ E-mail: m.kaiser@newcastle.ac.uk
\end{flushleft}

\section*{Abstract}
Complex networks have been characterised by their specific connectivity patterns (network motifs), but their building blocks can also be identified and described by node-motifs---a combination of local network features.
One technique to identify single node-motifs has been presented by Costa et al. (L. D. F. Costa, F. A. Rodrigues, C. C. Hilgetag, and M. Kaiser, Europhys. Lett., \textbf{87}, 1, 2009). Here, we first suggest improvements to the method including how its parameters can be determined automatically. 
Such automatic routines make high-throughput studies of many networks feasible. 
Second, the new routines are validated in different network-series.
Third, we provide an example of how the method can be used to analyse network time-series.
In conclusion, we provide a robust method for systematically discovering and classifying  characteristic nodes of a network.
In contrast to classical motif analysis, our approach can identify individual components (here: nodes) that are specific to a network.
Such special nodes, as hubs before, might be found to play critical roles in real-world networks.


\section*{Introduction}
Networks appear in a variety of real-world systems ranging from biology to engineering~\cite{Bornholdt2003, Boccaletti05}.
Examples include neural~\cite{Sporns2004, Kaiser2007rsta, Bullmore2009}, social~\cite{Lazer2009, Borgatti2009, Centola2010}, and computer networks~\cite{Albert1999diameter, Faloutsos1999power} to name but a few.
Networks have been used to study the emergence of cooperative behaviour~\cite{Nowak2006, Szabo2007, Perc2010}; to address epidemiological questions~\cite{Barthelemy2005, Funk2010} especially in scale free networks~\cite{PastorSatorras2001b, Meloni2009b}; and to investigate the causes of cascade effects~\cite{Watts2002, Buldyrev2010} for a more complete understanding of why networks differ in robustness against error and attack~\cite{Albert2000eaa, Kaiser2007}.
Attempts to classify network-topologies~\cite{Estrada2007} were accompanied by detailed studies of scale-free~\cite{Barabasi2009} and small-world networks~\cite{Watts1998, Schnettler2009}---properties that were identified in many real networks.
Additional to investigations of concrete structures, theoretical studies of random networks collected valuable information about large classes of networks~\cite{Albert2002, Newman2003, Newman2006}. 

Mapping complex systems to networks revealed that some nodes are remarkably different from other nodes of the same network. 
For instance, hubs, characterized by a high number of connections (a high node degree), often play a fundamental role in protein-protein interaction networks and their removal can be lethal for an organism~\cite{Jeong01_Nature, Rodrigues09}.
Hubs are similarly important for socio-economic systems, where defective hubs can cause cooperation to decline~\cite{Perc2009}.
Also, in engineered systems like the Internet, hubs are important to maintain the communication between autonomous systems~\cite{Albert2000eaa}.
These outlier nodes have been identified since the introduction of complex network theory, e.g. in the World Wide Web~\cite{Albert1999diameter} and the Internet~\cite{Faloutsos1999power}, but hubs are outliers only in terms of their degree; other network properties can also define special nodes. 
For instance, Internet topology has been shown decompose onion-like into different shells around a relatively small core network~\cite{Carmi2007}.
The closer a node's layer is to the core, the higher is the node's shell-index (coreness)~\cite{Seidman1983}.
Nodes with high coreness are not necessarily hubs, which one might suspect to be the most efficient spreaders of information.
Instead, the position of a node close to the network-core has more impact on successful dissemination than having a high degree~\cite{Kitsak2010}.
In networks where hubs are not present, as in most geographical networks, nodes whose neighbours are also connected to each other are special (high local clustering coefficient).
Further examples of outlier nodes can be found with different measures some of which examine more than the direct neighbourhood of a node~\cite{Albert2002, Costa2007a}, such that they specify rather global (network specific) than local (node specific) characteristics.
Global measures, such as characteristic path length or clustering coefficient~\cite{Watts1998}, summarise the whole network in a single value.
Local measurements, on the other hand, analyse each node or edge individually, yielding a more fine-grained picture of the network.
Nodes that express common features and outliers that are different can be identified with pattern recognition approaches, which group nodes of similar characteristics.
Corresponding techniques have been proposed recently~\cite{Newman07_PNAS, Wang08_NJP, Costa09_EPL} and revealed important network properties.
For example, in protein-protein interaction networks the relative number of outliers tends to decrease with the complexity of organism, i.e. proteins in more complex species show higher homogeneity in their interplay~\cite{Costa09_EPL}.
This demonstrates that, by considering multiple node-features jointly, pattern recognition based methods can point out exceptional network components.

Networks can describe complex systems whose interactivity between dynamical components changes over time.
Altered connections between the elements (represented by nodes) may in turn feed back on the dynamics, such that the dynamical process and the network topology evolve in an adaptive fashion~\cite{Gross2008}.
In the context of game theory, corresponding coevolution of behaviour and connectivity has been studied in socio-economic systems~\cite{Szabo2007, Perc2010}.
In complex scenarios like these, analysing a single network is often insufficient and several networks must be compared to gain insights.
Further examples for the need of network-comparisons are families of protein-protein interaction networks, brain connectivity networks in patient- and control-populations, or time-dependent (developing or declining) networks~\cite{Saavedra2008}.
Comparing such sets of networks requires consistent approaches, which are often non-trivial, because networks differ in size (number of nodes or edges) or they comprise a disjoint sets of nodes (some nodes occur in one network but not in others).
Direct comparisons between structures may thus be ruled out.
Based on outlier-detection as described above, we previously proposed motif-regions for which the relative frequencies of outliers falling into one of them yields a network specific fingerprint \cite{Costa_2009}.
Relating different networks to each other has thereby become as easy as comparing bar-graphs.
Nevertheless, although this methodology has been demonstrated to be suitable and accurate for outlier identification as well as for network comparisons, it suffers from several limitations, which we address in this paper.

Here, we describe a novel workflow for detecting characteristic single-node motifs and for using fingerprints for network comparison.
Improvements compared to the previous approach include (a)~automatic parameter determination, which facilitates high throughput analysis without user interaction, and (b)~replacing the k-means clustering algorithm with a deterministic method to simplify the workflow and to improve robustness of results.
In addition, we provide (c)~a validation of our method and (d)~an application to  networks where the topology changes over time (addition or deletion of nodes or edges).

\subsection*{Previous work}
The application of single measures to complex networks has revealed important insights in many cases. However, as Newman and Leicht recognised \cite{Newman07_PNAS}, detecting exceptions is limited to network features that are quantified by the measures in use.
Otherwise, if the chosen characteristics do not reflect the properties that are specific for a network or its components, important features remain unnoticed.

To solve this problem, two complementary approaches have been suggested. 
The first approach by Newman and Leicht groups nodes based on their connectivity without any further prior information \cite{Newman07_PNAS}. 
By fitting the parameters of a mixture model (using an expectation-maximization algorithm), each node is assigned a probability of belonging to any one group that has been identified. 
The probabilistic nature of this approach has the advantage that nodes that can not be unambiguously categorised are not crudely assigned to one particular group, but the conflict becomes evident, such that it can be dealt with.
The structure of networks can thereby be investigated without requiring any other parameter than the number of groups that are to be created.
This elegant method has been examined thoroughly and improvements to it have been suggested~\cite{Ramasco2008, Wang08_NJP}.

Analyses with focus on only one particular aspect of a network at a time might fail to detect irregularities or similarities in structure.
The second approach is to avoid single measures and to use a combination of multiple ones~\cite{Costa_2009}.
Instead of reducing network components down to one dimension, joint measures map it into a multi-dimensional feature space~\cite{Costa2007}; each vector-point in that space corresponds to a combination of node-characteristics and statistical methods are used to identify motif regions, such that each vertex falls into one of them: A node is either classified regular---showing features like the majority of nodes---or singular, i.e. its features deviate by following a particular single node-motif. 
The term motif refers to the concept of network motifs, i.e. patterns incorporating multiple nodes~\cite{Milo2002}.

Each of these two approaches to identify patterns in complex networks has its drawbacks and advantages.
The Newman and Leicht algorithm~(NLA) does not depend on one or few network measures, but it works on network links directly.
Networks are not restricted to undirected ones, but directed links and even weighted ones can be considered.
The NLA requires the number of node-groups to be specified; this is also true for the approach by Costa et al. [Beyond the Average (BtA)], where the number of motif regions needs to be chosen {\it a priori} \cite{Costa_2009}.
Unfortunately, for real-world networks this number is often unknown.
The BtA-workflow requires two additional parameters to control which nodes will constitute individual motif regions.
Both methods differ in their output, as BtA not only provides a grouping of nodes, but also a network-fingerprint, which can be used to compare networks from different domains.
Most importantly, however, is the conceptional distinction between NLA and BtA, as they rely on local edge connectivity and local node measures, respectively.
BtA will fail to pinpoint features of the network, if the chosen set of measures can not formulate a corresponding motif.
Similarly NLA can fail, as it only takes into account direct connections between nodes: NLA does not consider how the neighbours of a node are connected, for example, but BtA can deal with such information (by evaluating the local clustering coefficient).
Indeed, the extensibility concerning features to assess is the biggest advantage of BtA; (un-)directed and weighted links can be processed likewise and in spatial networks the location of nodes can be taken into account.
In conclusion, NLA is readily applicable to a broad variety of network domains; however, considering direct connections only is a weakness.
BtA can be nicely adopted to these cases, but care has to be taken at all times to ensure the set of measures is diverse enough to cover as many patterns that might occur in networks as possible. 

\begin{figure*}[!ht]
    \begin{center}
    \includegraphics[width=0.8\textwidth]{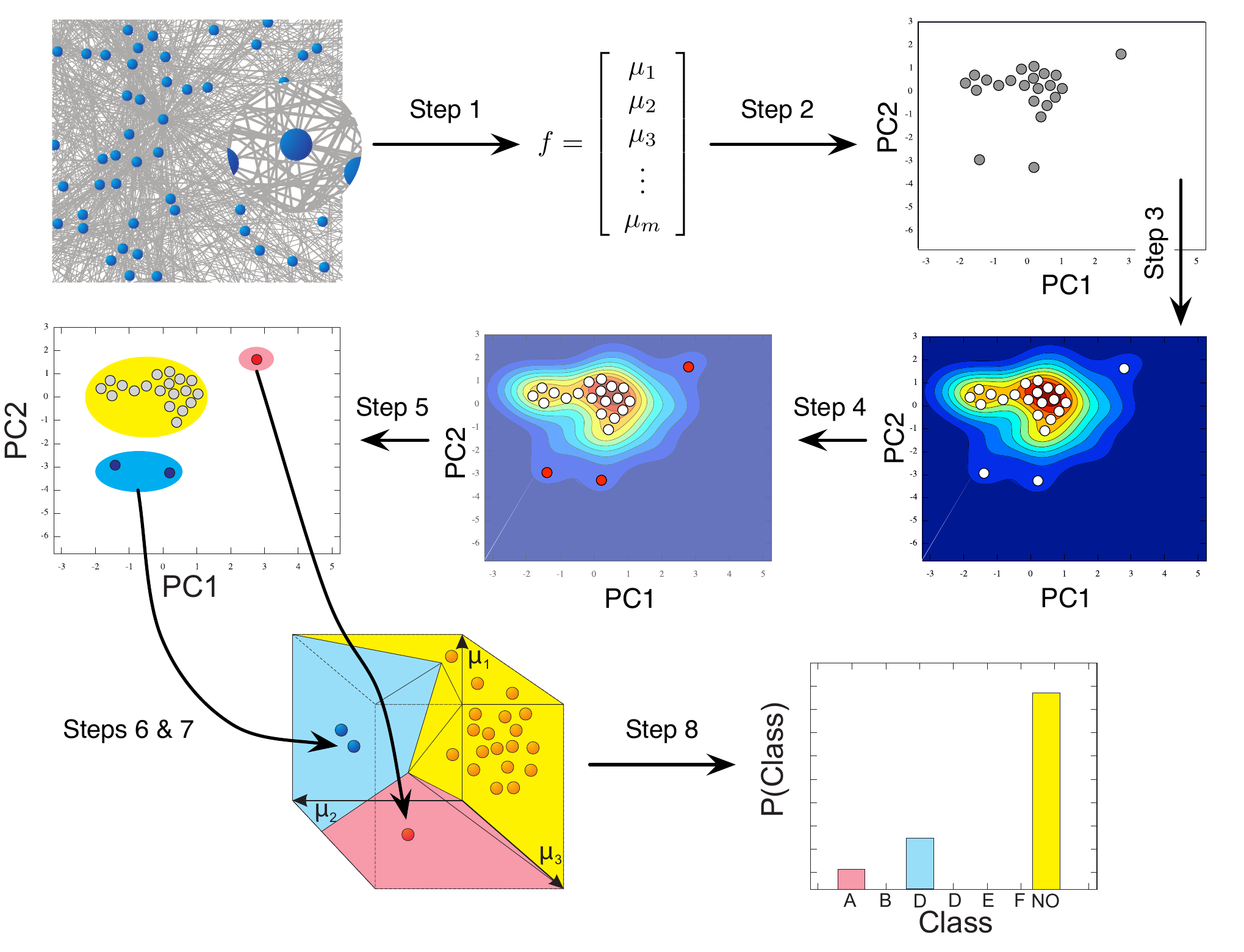}
    \end{center}
    \caption{ 
            {\bf Analysis work-flow to identify global singular nodes from local features~\cite{Costa_2009}.}
            Step~1: Choose set of local measures to characterise network nodes~\cite{Costa2007a}. Calculate local measurements for all nodes in the network (\textit{feature vectors}).
	    Step~2: Map each node's feature vector to lower dimensional space using principal component analysis (PCA plane)~\cite[Chapter~8]{Johnson2007}.
	    Step~3: Estimate each node's probability using the Parzen window approach (PDF)~\cite[Chapter~4.3]{Parzen1962, Duda2001}.
	    Step~4: Query PDF to identify least probable nodes (\textit{singular nodes}).
	    Step~5: Cluster singular nodes in PCA plane using k-means (\textit{motif groups})~\cite[Chapter~20.1]{Mackay2003}.
	    Step~6: Determine Voronoi cells for grouped nodes using a modified Mahalanobis distance (\textit{potential motif regions})~\cite{Mahalanobis1936}.
	    Step~7: Join potential motif regions that are close to each other (\textit{motif regions}).
	    Step~8: Calculate relative frequencies of nodes falling into motif-regions~(A--F) or non-motif region~(NO) (\textit{fingerprint}).
        }
\label{fig_workflow_illustration}
\end{figure*}

In the next section we suggest several improvements to the BtA-workflow (Fig.~\ref{fig_workflow_illustration}), which can be sketched as follows:
Initially, multiple local network measures are applied to each node, which yields a multi-dimensional characterisation in form of a {\em feature vector}.
Correlation between different measures is accounted for by principal component analysis~(PCA), which is used to map feature vectors of all nodes to two dimensional space~\cite[Chapter~8]{Johnson2007}.
Next, nodes are assigned probabilities in order to distinguish nodes with common and rare features.
The required probability density function~(PDF) is gained by smoothing over points in the two dimensional PCA-plane (Parzen window approach~\cite[Chapter~4.3]{Parzen1962, Duda2001}).
Now, the least probable nodes, i.e those with uncommon features, can be identified from the PDF.
These {\em singular nodes} are then clustered in order to distinguish different {\em motif-groups}.
Each of the two dimensional motif-groups corresponds to a higher dimensional {\em motif-region} into which the feature vectors split up and the distribution of feature vectors among the different motif-regions is the {\em fingerprint} of the network.
Apart from the initial decision on which measures to use, the user needs to choose the bandwidth of the smoothing kernel, the number of singular nodes~$w$, and the number of motif-groups~$k$, respectively (steps~3, 4, and~5 in Fig.~\ref{fig_workflow_illustration}).
Additionally, when comparing multiple networks, a limit must be specified below which motif regions are considered too close to each other to constitute different motifs (join threshold; Step~7).
So far, these settings had to be chosen manually, but here we suggest how to determine all three parameters (bandwidth, $w$, and~$k$) automatically.
The last setting (join threshold), however, is not considered for automation:
So far we could not identify a procedure that yields results as good as manual selection by the user.
We thus concentrated our efforts on the parameters that need to be set for every network (bandwidth, $w$, and~$k$), such that high-throughput applications become possible.
Automating the setting of the main parameters is thus of higher benefit than for the threshold that determines Voronoi cells to be joined; this needs to be chosen only once, when all networks are compared to each other at the end.

\section*{Results}
In this paper we propose how to choose all relevant parameters of the BtA-workflow automatically (see Methods section), which allows for the analysis of many networks without the need for human interaction.
The only remaining limiting factor for high throughput analyses are the computational costs of the analysis, which predominantly depend on the measures that are chosen to characterise each node.
Using implementations of common local measures (see supporting information), the estimated run-time scales linearly to cubic with network size (Fig.~S1).
Costs are thus comparatively cheap considering methods that identify specific connectivity patterns by counting occurrences of particular sub-graphs (e.g.~\cite{Wasserman1994, Kuramochi2001, Milo2002, Kashtan2004, Middendorf2005, Bordino2008}); such motif-counts also scale at least linearly in network size, but they show exponentially growing costs as the size of the motif-pattern increases~\cite{Kashtan2004}.
In practice this often means that counts can not be determined for patterns involving 10~nodes or more~\cite{Ribeiro2009}, which renders some domains computationally intractable for this approach, but eventually not for BtA.
However, before processing huge networks or many different structures with BtA, we first need to verify that parameters are indeed chosen adequately, which is confirmed in the next section.

\subsection*{Method Verification}
\begin{figure}[!ht]
    \begin{center}
    \includegraphics[width=0.4\textwidth]{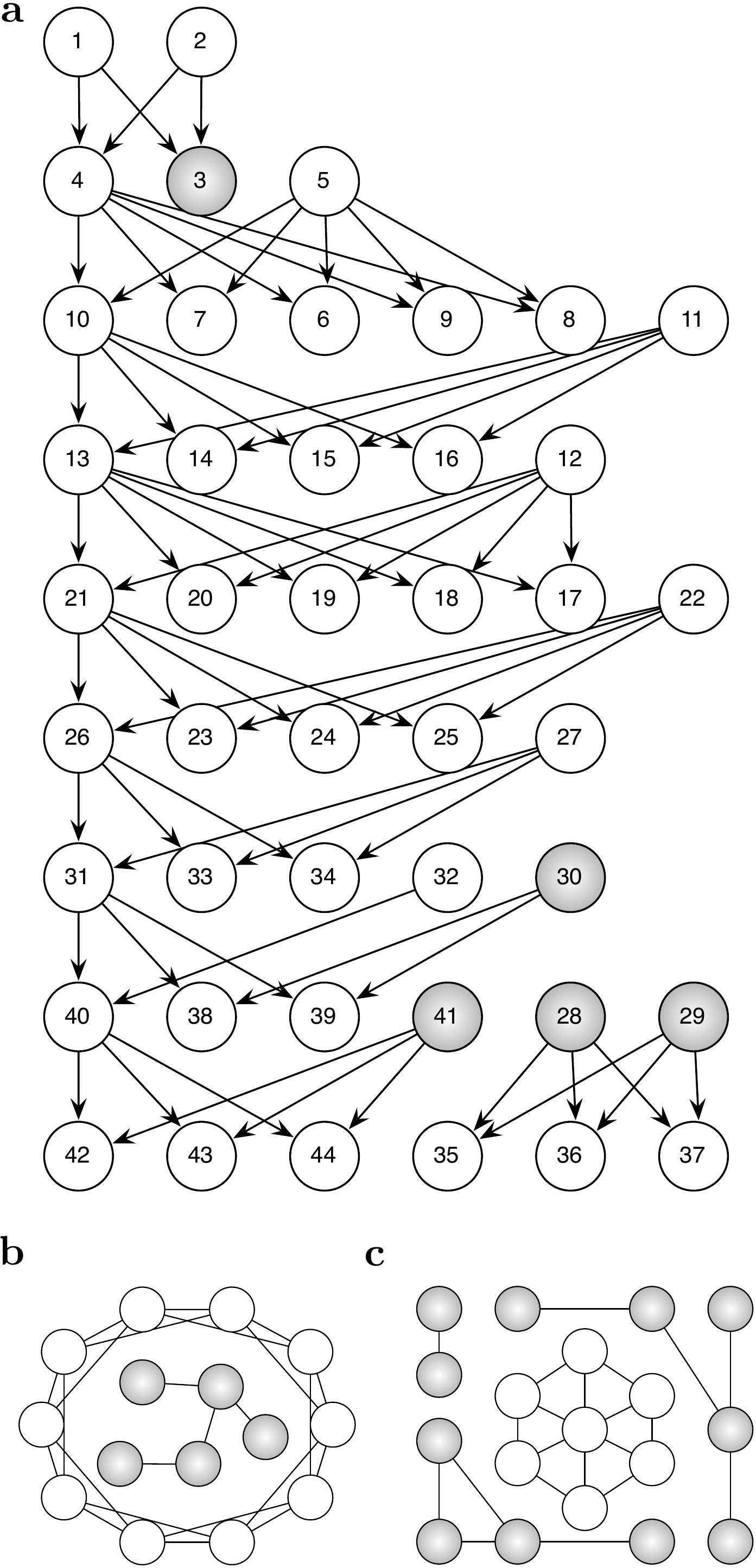}
    \end{center}
    \caption{ 
	{\bf Network types used for testing BtA:}
    	\textbf{a}~Network derived from {\em The Simpsons} family-tree~\cite{Groening2005}. 
    		Nodes in very regular parts of the network were identified singular (shaded grey) because of two characteristics: Their neighbours' degrees are comparatively low and show no variation (values $r$ and $cv$ significantly below average).
    	\textbf{b}~Schematic of large regular ring lattice combined with a minor ER-random component (shaded grey).
    	\textbf{c}~A small regular structure (white nodes) embedded into a large random network (ER, BA, or WS model).
        }
\label{fig_verification}
\end{figure}

The first validation is on a network that is small enough to confirm BtA-results by eye:
We use a family-tree from {\em The Simpsons}~\cite{Groening2005} to create a network with nodes representing characters and directed links pointing to their offspring (Fig.~\ref{fig_verification}a).
Nodes that have a sparsely connected and homogeneous neighbourhood are suitably highlighted as outliers by BtA.

With these reassuring results from a single network, we proceed by testing BtA on whole series: We generate structures with both regular components and exceptional ones, which BtA has to identify.
In our first series we compose networks of two components: a regular ring lattice and a smaller Erd\H{o}s-R\'enyi~(ER)~\cite{Erdos1959} random network (Fig.~\ref{fig_verification}b).
While the ring lattice remains unchanged, the size of the random module increases throughout the series, such that its proportion of the full network grows gradually\footnote{
The ring lattice is comprised of 100~nodes, each of which is connected to its four closest neighbours (Fig.~\ref{fig_verification}b).
ER-random networks ($n = 1, \ldots, 50$ nodes) have an average edge-density of 25\%.
}.
Composed networks are analysed with BtA:
Of all outlier-nodes less than 2\% are missed while over 96\% are classified correctly, if the random component contributes less than 25\% of nodes to the network.
Beyond that limit, the number of nodes in the random-part does no longer match the number of identified outliers~$w$.
But this does not imply a mis-classification by BtA:
The larger a random network, the more likely it is that a few nodes are connected regularly (or close to that).
Quantifying these nodes with local network measures yields the same values as (or similar to) those of the ring lattice, which is why it would be incorrect to consider them singular.
Additional to regular connection patterns in large random networks, other local motifs can also be frequent enough, such that they constitute a common rather than an exceptional feature of the network.
Thus, network components that seem clearly separable at first may actually be very similar or---although intended to form outliers---they may contain common elements, due to random effects.
Together this explains the observed deviations in numbers of outliers for growing ER-components in this test-series.

Finally, we reverse the nature of the networks: The major component is set to a random network [ER, Barab\'{a}si and Albert~(BA)~\cite{Barabasi1999}, or Watts-Strogatz~(WS)~\cite{Watts1998} model] in which we embed a small, but highly regular structure (Fig.~\ref{fig_verification}c).
The inserted structure was chosen, such that its nodes are highly clustered (both on level~1 and~2); the six outer nodes further show significant variability in their neighbours' degrees.
These characteristics are rarely observed in our random networks, which is why BtA should identify these nodes (alongside with other outliers that might emerge).
We confirm this in a series of networks with varying sparseness\footnote{%
Random networks ($n = 100$ nodes) are generated according to the ER, BA, and WS model; edge-density is gradually increased from 1\% to 50\% (step-size 1\%). 
The regular structure (7~nodes) illustrated in Fig.~\ref{fig_verification}c is added to each random network before BtA is applied.
}:
The 6 outer nodes of the regular structure are classified singular in over 97\% of all networks.
Additionally, the inner node (with less extreme features) is regarded uncommon in 81\% of all cases.

In conclusion, the automatic parameter determination gives very satisfying results, which yield confidence in BtA's ability to identify outliers in complex networks autonomously.

\subsection*{Network Time-Series: A Small-World Emerging}
Large complex networks are challenging to analyse; time-series of such are even more so.
We attempt to approach this challenge by first condensing networks to a compact representation---mapping a series of changing structures to a uniform representation benefits the identification of trends and changes of such.
Therefore, all networks have to be characterised, which we do using single node-motifs.
These are identified with BtA using six common local measures:
(1)~the normalised average degree~$r$,
(2)~the coefficient of variation of the degrees of the immediate neighbours of a node~$cv$,
(3)~the clustering coefficient~$cc$~\cite{Watts1998, Kaiser2007njp},
(4)~the locality index~$loc$,
(5)~the hierarchical clustering coefficient of level two~$cc_2$~\cite{Costa2006},
and (6)~the normalised node degree~$K$.
(For definitions of these measures see Methods section.)
Next, we describe the time-series of 600 networks and the results found with BtA.

Similar to random graphs, small-world networks have a small characteristic path length, but at the same time they exhibit a high degree of clustering, as regular ring lattices, for example.
It has been discovered early that the combination of short paths plus grouping is inherent to social networks; a phenomenon that became known as six degrees of separation~\cite{Milgram1967, Kochen1989, Guare1994}.
Today it is known that small-world networks can be found in many other domains (e.g.~\cite{Albert2002, Newman2003, Boccaletti05, Newman2006}).
We thus created a network-time series in which structures gradually change from a completely regular ring lattice to a small-world network (see Methods section, Fig.~S2).

\begin{figure*}[!ht]
    \begin{center}
    \includegraphics[width=0.8\textwidth]{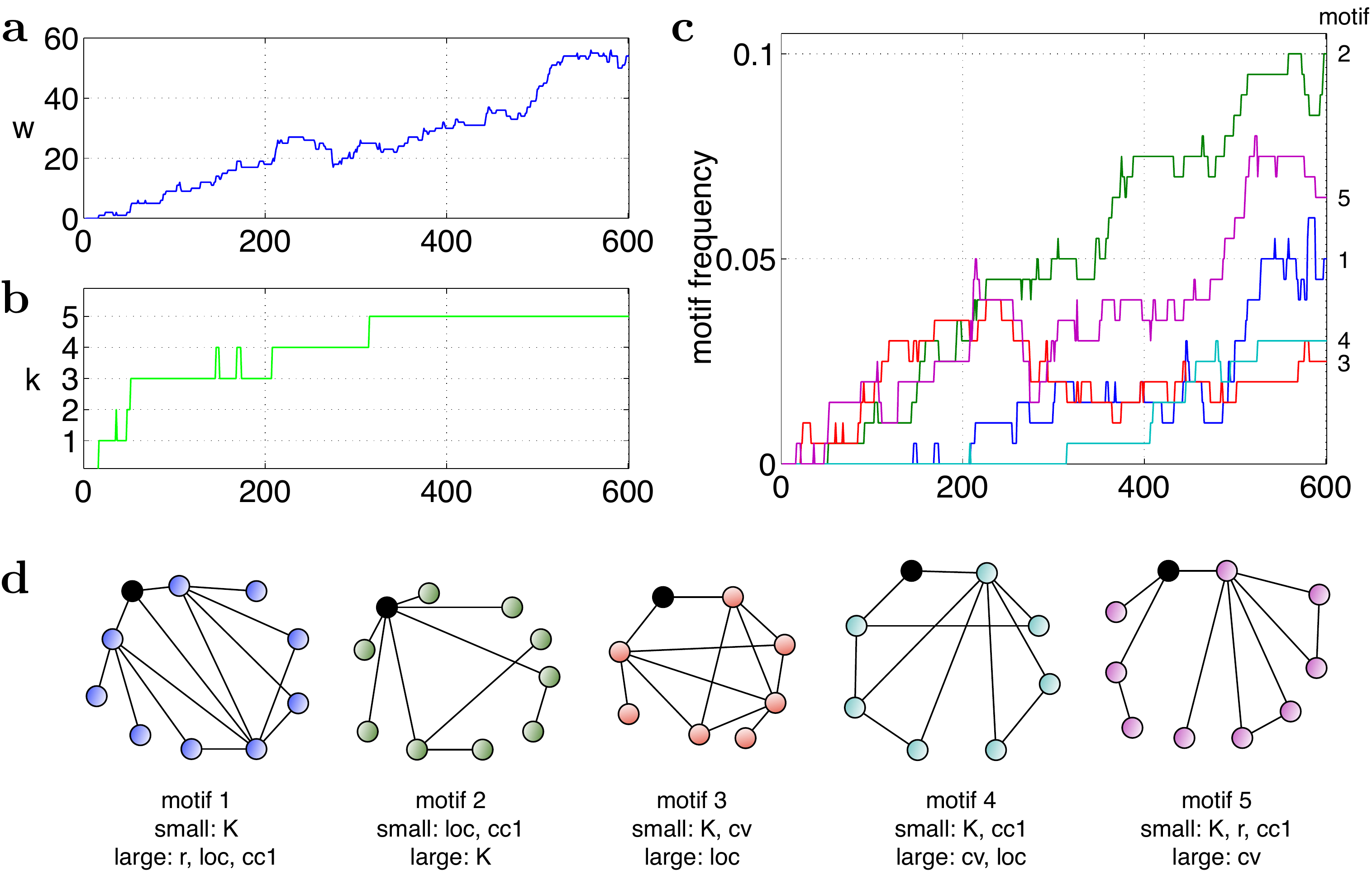}
    \end{center}
    \caption{
    	{\bf Single node-motifs in emerging small-world network~(Fig.~S2).}
    		Vertical axes in subfigures~a--c correspond to number of outlier nodes~$w$, number of single node-motifs~$k$, and their frequencies, respectively.
    		\textbf{a}~Number of identified outliers~$w$ rising from 0 to~54.
			\textbf{b}~Diversity of node-motifs~$k$ quickly rising during the 1\textsuperscript{st} re-wiring round; less increase during 2\textsuperscript{nd} round; and stable during the~3\textsuperscript{rd}.
			\textbf{c}~Proportions of nodes expressing identified motifs ({\it motif frequencies}). 
				Nodes classified regular not shown. 
			\textbf{d}~Schematics of identified single node-motifs and their distinguishing characteristics.
        }
\label{fig_rewiring_results}
\end{figure*}

In total we identified 5 single node-motifs, which differ in characteristics, frequency, and time of emergence (Fig.~\ref{fig_rewiring_results}):
A node according to motif~1 has relatively few connections in contrast to its well connected neighbourhood.
Different from that, nodes corresponding to motif~2 are signified by many connections to a rather sparsely connected neighbourhood.
Motif~3-nodes have relatively few connections and nodes in their neighbourhood are similar in number of links and corresponding targets.
Motif~4 describes rarely connected nodes whose neighbours have a diverse number of connections; but instead of being linked between each other, neighbours share other common targets.
The final motif~5 can be best characterised by its relation to the rest of the network, which shows a higher degree of connectivity than any node involved in the motif.
Neighbours of the motif-node further vary in their number of connections and do not link to each other.
Motifs~2, 3 and~5 appear right from the beginning of the rewiring process; motifs~2 and~5 gradually become more common over time, whereas~3 levels out after a transient peak.
The remaining motifs~1 and especially~4 only become apparent at later stages towards which both become more frequent.
Together, BtA reveals the increasing irregularity in network structure and it also provides details on the characteristic connectivity patterns at different times.
Both would be valuable information if real networks were analysed; here, with precise knowledge about the network-changing process, the temporally dependent motif expression levels yield another validation of the technique (detailed discussion in supporting information).

Overall, results are very satisfying and we are confident that BtA could be successfully applied to real networks using the automatic parameter determination.

\section*{Discussion}
In this paper we presented a method to detect single node-motifs automatically. 
The main parameters of the previous routine~\cite{Costa_2009}---the smoothing kernel bandwidth plus the number of singular nodes and motif groups---are now selected based on the data.
We further proposed a deterministic replacement for the k-means algorithm, which is used to form the different motif-groups.
In contrast to k-means, our alternative approach can determine the number of motifs itself and due to the lack of random elements, clustering results are robust over multiple repetitions.

Despite our improvements to BtA certain issues and room for further advancements remain.
For example, reducing feature vectors in dimension inevitably leads to a loss of information, but which has to be kept withing reasonable bounds.
In other words, although 6-dimensional feature vectors were suitably represented in the 2-dimensional plane so far \cite{Costa_2009}, different networks may require the use of more than just the first 2 principal components in order to ensure that network characteristics are represented properly.
Thus, if the chosen number of principal components does not account for at least 80\% of the variance, their number should be increased ({\em Kaiser's rule}).
The degree to which feature vectors can be reduced thereby depends on the correlation between measured values, which is specific to the analysed network.

In cases where feature vectors can not be suitably represented in 2 dimensions, their display becomes more complicated and verifying a good fit of the estimated probability density function (PDF) is challenging.
However, a good PDF estimate is needed in the BtA workflow to determine outlier nodes.
Problems that might arise in these situations could possibly be circumvented by a major change to the workflow: 
The use of PCA to compact information offers the possibility to replace both the PDF estimation and the subsequent outlier selection with a more direct and non-parametric standard technique, which is {\em Hotelling's~$T^2$} (a generalisation of {\em Student's t-statistic}).
This modification would allow to identify outliers without the need to estimate a PDF, but the exploration of the resulting workflow will be addressed in another publication. 

Considering the BtA workflow as presented in this paper, the technique can be easily adapted by including different local network measures in the analysis.
Measures that take spacial aspects of the network into account, for instance, or those including link-weights can increase quality of the analysis.
Finally, interest might not only lie on motifs formed by outlier nodes, but on all single node-motifs occurring in the network.
In this case regular and singular nodes are not distinguished, but all of them have to be included in the network fingerprint.

BtA-fingerprinting of many networks has so far been prevented by the need to choose parameters during the analysis manually.
With the improvements presented in this paper, however, it is now possible to process large numbers of networks fully unsupervised. 
Identified outliers are characteristic nodes that can provide a fingerprint of a network; fingerprinting networks from numerous domains allows easy characterisation and comparisons.
As already demonstrated~\cite{Costa_2009}, such studies can reveal important characteristics and differences between network domains.
Additionally, the example on an emerging small-world network in this paper showed that BtA can also be used to analyse time-series of networks.

To encourage the use of the BtA methodology by other researchers, we provide our implementation of the workflow including the automatic parameter determination for download (\url{http://www.biological-networks.org/}).
Two versions of the code exist:
The first one requires Matlab (Mathworks Inc, Natick, USA) and allows the user to apply the workflow using a graphical user interface (Fig.~S3).
The other one is a command line utility that either requires Matlab or the free alternative Octave~\cite{Eaton2002} and it can be easily used to batch process many networks.

In conclusion, we provide a robust method for systematically discovering and classifying  characteristic nodes of a network.
The distribution of node-classes results in a fingerprint, which in turn can give a classification of whole networks, as for network motifs of multiple nodes~\cite{Milo2004}.
In contrast to classical motif analysis, our approach can identify the individual components that are specific to a network.
Such special nodes, as hubs before, might be found to play critical roles in real-world networks.

\section*{Methods}

\subsubsection*{Local Network Measures} 
Network nodes were characterised with six common local measures whose definitions are given in the following.
Therefore, let~$A=(a_{ij})$ denote the adjacency matrix of the network, i.e. $a_{ij}=1$, if a link from node~$i$ to node~$j$ exists, and otherwise~$a_{ij}=0$.
Row- and column-sums of~$A$ correspond to the {\em in-} and {\em out-degrees} of nodes, respectively.
In undirected networks, in- and out-degree are equal and either of them can be used as a node's {\em degree}.
If links are directed, the degree is the sum of in- and out-degree.
Dividing a node's degree by the number of all links in the network yields the {\em normalised node degree}~$K$.
The {\em normalised average degree}~$r_i$ of a node~$i$ is the average over all its neighbours' degrees.
(Nodes that are directly linked to node~$i$ are called {\em neighbours}.)
Likewise, the {\em coefficient of variation}~$cv$ of the degrees of the immediate neighbours of a node can be calculated.
The neighbours' connectivity with each other is quantified by the {\em clustering coefficient}~$cc_i$, which is the proportion of existing connections between node~$i$'s neighbours to the number of all possible links between them~\cite{Watts1998, Kaiser2007njp}.
The clustering coefficient thus reflects the relative number of triangle-shaped paths a node has---a concept that is extended to connections between neighbours' neighbours (further away node node~$i$) by the {\em hierarchical clustering coefficient} of level two~$cc_2$~\cite{Costa2006}.
Whereas the cluster coefficients quantify connectivity within a node's neighbourhood, the {\em locality index}~$loc_i$, which is based on the matching index (e.g.~\cite{Kaiser2004}), is the fraction of neighbours' links that connect to the same node (not necessarily a neighbour of node~$i$).
Further details and measures can be found in the literature~\cite{Albert2002, Newman2003, Newman2006, Costa2007a}.\\

\noindent
In the following sections we describe how appropriate settings for the parameters of the BtA-workflow can be found automatically.
Kernel-bandwidth, the number of singular nodes~$w$, and the number of motif regions~$k$ are discussed separately below.

\subsubsection*{Kernel-Bandwidth} 
In step~3 of the workflow (Fig.~\ref{fig_workflow_illustration}), the Parzen window approach is used to estimate a probability density function (PDF) over all nodes~\cite[Chapter~4.3]{Parzen1962, Duda2001}.
This is achieved by smoothing the overall arrangement of reduced feature vectors, which were obtained using principal component analysis~(PCA)~\cite[Chapter~8]{Johnson2007} in the previous step~2.
The dimensions of the smoothing kernel, i.e. the width and breadth of the Gaussian function~$\mathcal{N}_2(\mu,\, \Sigma)$ can be controlled through its covariance matrix~$\Sigma = (\sigma_{ij})$. 
(Mean vectors~$\mu$ are fixed to equal the data-points.)
The original publication made use of the fact that the absolute covariance values ($\forall k:\, \sigma_{kk} = 0.05$) do not matter for the estimated PDF.
However, their values relative to each other do matter and we therefore scale them according to the standard deviation along each principal component (PC) axis.
Variability-based re-shaping of the kernel function improves the overall fit of the PDF to the points (Fig.~S4).
A further refinement would be to tilt the Gaussian in order to account for correlation between axes (Fig.~S5); however, the PCs are expected to show weak correlation only, which is why we chose un-tilted kernels (for which the covariance matrix~$\Sigma$ is zero except for the variances on the diagonal).

\subsubsection*{Number of Singular Nodes~$w$} 
After assigning probabilities to all nodes (Step~3), nodes with an exceptionally low probability come into focus:
These outliers correspond to points in the PCA-plane that are spatially separated from larger clusters; and this separation corresponds to abnormalities of measured features.
Due to their uncommon characteristics, these nodes are considered singular. 
For humans it is usually straightforward to identify these non-regular nodes, if interactive visual aids are provided; we therefore implemented a graphical user interface for the whole workflow (Fig.~S3).
In the following, however, we discuss how the number of singular nodes~$w$ can be adjusted without interaction.

To determine singular nodes, automated methods can query the PDF that has been estimated earlier (Step~3).
For example, for a fixed number~$w$ of singular nodes, the $w$~least probable ones can be selected easily.
Alternatively, a probability cut-off can be set, e.g. at 1\% or 5\%, to separate nodes into regular and singular ones.
Both these simple methods involve constants, but which have to be chosen depending on network size to yield sensible results\footnote{Choosing one fixed number of singular nodes~$w$ for differently sized networks can render the majority of nodes non-regular in comparatively small networks; vice versa, $w$~may be too small compared to the number of exceptional nodes in large networks. A fixed probability cut-off does not circumvent this problem, because the nodes' absolute probability values are dependent on network size.}.
In the following, we therefore propose a flexible probability-threshold: The cut-off does not occur at a fixed pre-defined level, but where it yields the best separation between singular and regular nodes.

A necessary condition for a node being considered singular is a sufficiently low probability compared to other nodes.
Additionally, it is desirable that singular nodes appear somewhat separated from the regular ones, which renders their classification non-arbitrary.
We therefore suggest to set the borderline between regular and singular nodes where the steepest increase in probability among the low probability nodes appears.
Nodes with a probability below mean~$\bar{p}$ minus one standard deviation~$\sigma(p)$ of all nodes' probabilities~$ p = (p_k)_{k=1,\ldots,n} $ are potentially singular.
Given that the probabilities $ p = (p_k)_{k=1,\ldots,n} $ are sorted increasingly, the number of singular nodes~$w$ is then chosen as
\begin{equation} \label{eq_w}
	w = \arg\max_{k \ : \ p_k < \ \bar{p} - \sigma(p) }{ p_{k+1} - p_k } \ ,
\end{equation}
or $ w = 0 $, if probabilities undershoot the mean only minimally (i.e. $ \nexists \ k : p_k < \bar{p} - \sigma(p) $).

\subsubsection*{Number of Motif Groups~$k$}
Once nodes are classified as either regular or singular (Step~4), clusters of singular nodes (\textit{motif-groups}) are identified using $k$-means~\cite[Chapter~20.1]{Mackay2003}.
The $k$-means clustering algorithm requires the number of clusters~$k$ to be chosen {\it a priori}; the actual procedure then determines $k$~centroids and assigns each node to the closest one of them.
Choosing $k$ too low results in clustering errors, because multiple motif-groups are falsely considered as one.
Conversely, too many clusters split motif-groups into non-existing sub-groups.
Determining a suitable~$k$ is thus crucial for automating the workflow and we come back to this issue later.
Even if $k$ is chosen adequately, clustering results are not guaranteed to be satisfactory when using k-means:
The algorithm initially chooses the cluster-centroids at random, but their actual distribution impacts on the quality of clustering results~\cite{Jain1999}.
Attempts to optimise the centroid initialisation have been made (e.g. the $k$++-algorithm~\cite{Arthur2007}), but random effects still remain;
we therefore suggest a deterministic replacement for $k$-means.

\begin{figure}[!ht]
    \begin{center}
    \includegraphics[width=0.48\textwidth]{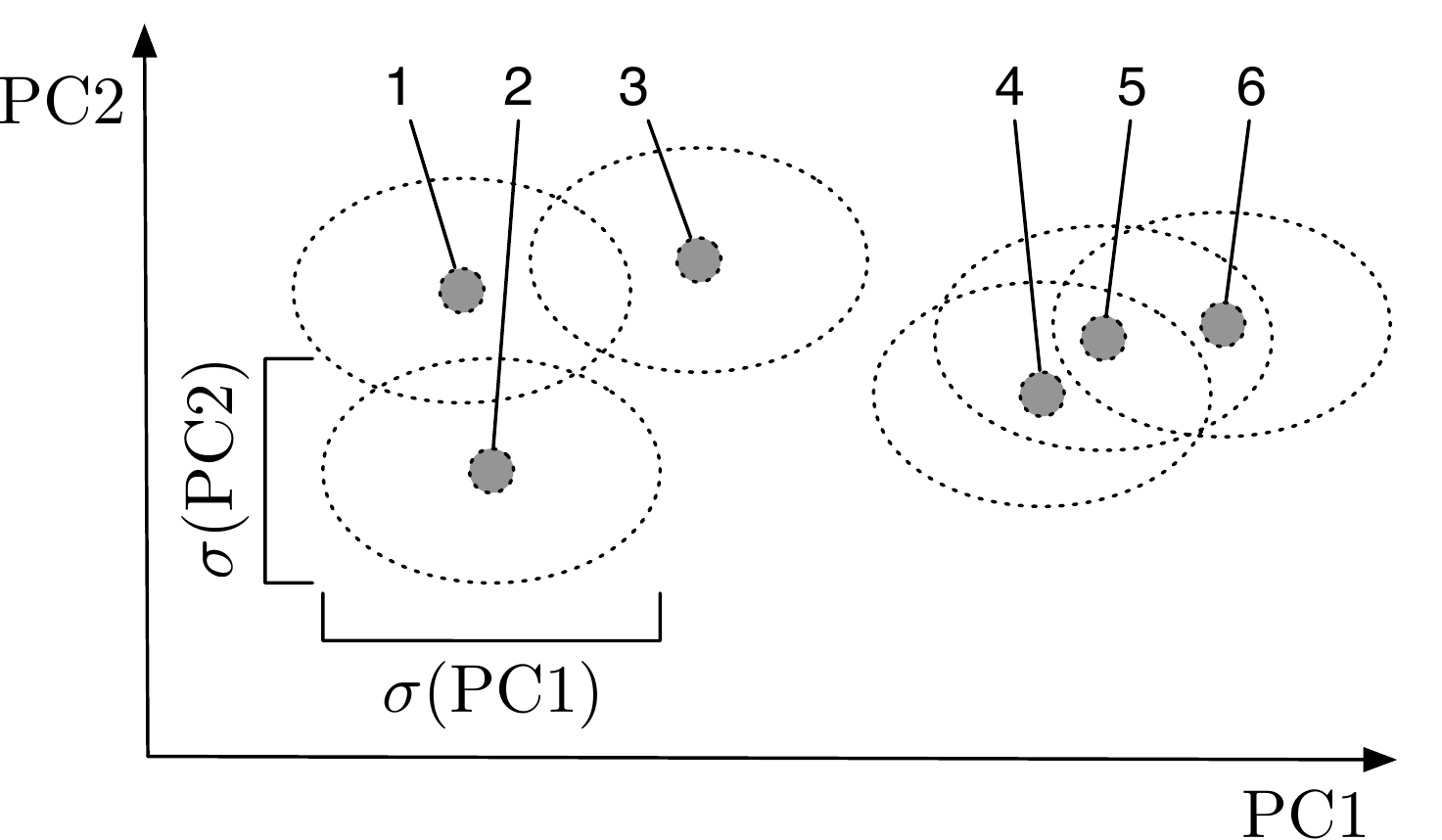}
    \end{center}
    \caption{
	 {\bf Example of 2 clusters (left, right) with 3 points each (1--3, 4--6).}
         Ellipses are centred on each point with dimensions corresponding to standard deviations~$\sigma$ along PC-axes.
         A set of points is considered a \textit{clique}, if the area of all their ellipses is connected (e.g. \{1\}, \{1, 2\}, or \{1, 2, 3\}; but not $\{2, 3\}$).
         A maximal clique is called a \textit{cluster} (i.e. \{1,2,3\} or \{4,5,6\}) and is used to define a distinct motif-group.
         }
\label{fig_overlap_grouping}
\end{figure}

Optimal groupings of singular nodes consider well separated nodes to be in different clusters, whereas relatively close ones are grouped together.
The standard deviations along each PC-axis can serve as a threshold for \textit{closeness} and we consider each of the singular nodes to occupy a certain volume in the PCA-plane, i.e. an ellipse-shaped area centred on it.
All ellipses have the same dimensions, which equal the standard deviations along the two axes.
Nodes are then assigned to the same motif-group if all their ellipses constitute a connected area (Fig.~\ref{fig_overlap_grouping}).
Practically, this idea can be implemented in 3~steps:
\begin{enumerate}
    \item Similar to an adjacency matrix, create a binary \textit{overlap-matrix}~$ O=(o_{ij}) $ in which nodes are connected if their ellipses overlap; otherwise they are not.
	For two nodes~$i$ and~$j$ let $x = (x_1, x_2)$ and $y = (y_1, y_2)$ denote their corresponding points on the PCA-plane, i.e. the centres of their ellipses with dimensions~$\sigma_1$ and $\sigma_2$.
	Using the rescaled centres $c_x = (x_1/\sigma_1, x_2/\sigma_2)$ and $c_y = (y_1/\sigma_1, y_2/\sigma_2)$ the entry of the overlap-matrix is defined by
    \begin{equation}
        o_{ij} = \begin{cases}
            1 , & d_2(c_x, c_y) < 1 \ , \\
            0 , & \text{otherwise} \ ,
        \end{cases}
    \end{equation}
    where $d_2(\cdot, \cdot )$ is the Euclidean distance.
    \item Determine a corresponding \textit{clique-matrix}~$ C = (c_{ij}) $ that specifies whether a path---a connected area of ellipses---between any two nodes exists or not.
    Paths or cliques can be determined through powers $ O^k = \left(o_{ij}^{(k)}\right) $ of the overlap-matrix~$ O $ via 
    \begin{equation}
        c_{ij} = \begin{cases}
            1 , & \exists \ k \in \mathbb{N} : o_{ij}^{(k)} > 0\ , \\
            0 , & \text{otherwise} \ .
        \end{cases}
    \end{equation}
    \item Colour all cliques differently, which finally yields the motif-groups.
\end{enumerate}
Note that this procedure has no parameter controlling the number of motif-groups, but these are identified automatically.
Instead of using this method to actually group nodes it might also serve as a pre-processing step in order to determine the number of clusters~$k$ for k-means.
The drawback of this simple approach is that long elongated clusters can result when nodes are widely distributed, but connected by a chain of nodes that are just less than one standard deviation apart from each other.
However, we have not observed such formation in practical applications.

\subsection*{Generation of Small-World Networks}
The prevalence of small-world networks has risen questions about their generating mechanisms and different explanatory models have been proposed~\cite{Watts1998, Ozik2004}.
We use one of them here in order to generate a series of networks:
Watts and Strogatz described a rewiring procedure by which a regular ring-lattice is randomly rewired by which it becomes a small-world network \cite{Watts1998}.
This is a step-wise process, which allows to sample a network at each intermediate stage.
Starting with a completely regular structure, over time, networks become increasingly perturbed (Fig.~S2).
In total, we sampled 600 networks (\`a~200~nodes), which were then analysed with BtA, to determine the single node-motifs that evolve over time.

\section*{Acknowledgments}
Marcus Kaiser and Christoph Echtermeyer were supported by EPSRC (EP/G03950X/1) and the CARMEN e-science project (\url{http://www.carmen.org.uk}) funded by EPSRC (EP/E002331/1). Marcus Kaiser also acknowledges support by the WCU program through the National Research Foundation of Korea funded by the Ministry of Education, Science and Technology (R32-10142).
Luciano da F. Costa is grateful to CNPq (301303/06-1) and FAPESP (05/00587-5) for financial support.
Francisco A. Rodrigues is grateful to FAPESP (2007/50633-9) for sponsorship.

\bibliography{p_citations}

\section*{Figures}
%
%
\newpage

\newpage
\newpage
%
%


\end{document}


\maketitle
\textsuperscript{1}{School of Computing Science, Claremont Tower, Newcastle University, Newcastle-upon-Tyne NE1 7RU, UK}

\textsuperscript{2}{Instituto de F\'isica de S\~ao Carlos,
Universidade de S\~{a}o Paulo, S\~{a}o Carlos, PO Box 369, 13560-970 S\~{a}o Carlos, S\~{a}o Paulo, Brazil}

\textsuperscript{3}{Departamento de Matem\'{a}tica Aplicada e Estat\'{i}stica, Instituto
de Ci\^{e}ncias Matem\'{a}ticas e de Computa\c{c}\~{a}o,
Universidade de S\~{a}o Paulo, S\~{a}o Carlos, PO Box 668, 13560-970 S\~{a}o Carlos, S\~{a}o Paulo, Brazil}

\textsuperscript{4}{Institute of Neuroscience, The Medical School, Framlington Place, Newcastle University, Newcastle-upon-Tyne NE2 4HH, UK}

\textsuperscript{5}{Department of Brain and Cognitive Sciences, Seoul National University, Seoul 151-746, Korea}

\textsuperscript{*}Corresponding author. Electronic address: m.kaiser@newcastle.ac.uk


\tableofcontents
\listoffigures

\newpage
\section{Supplementary Figures}
\begin{figure*}[!h]
    \begin{center}
    \includegraphics[width=0.9\textwidth]{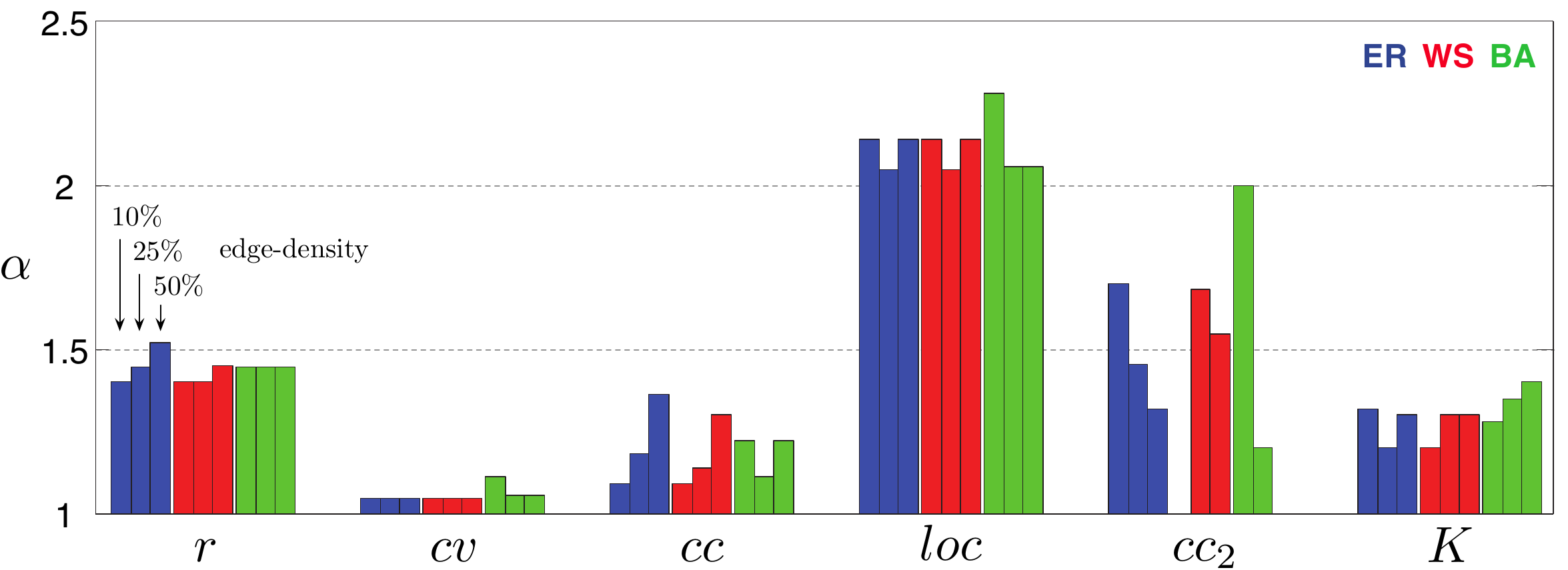}
    \caption[Run-time complexity of local network measures]{ \label{fig_complexity_comparison}
        Run-time complexity~$O(n^\alpha)$ of the used local measures ($r$, $cv$, $cc$, $loc$, $cc_2$, $K$) increases polynomially in network size~$n$ (average value for 100 networks). 
        Growth determining exponent~$\alpha$ depends on edge-density (10\%, 25\%, 50\%) and to lesser extent on network model (ER, WS, BA).
    }
    \end{center}
\end{figure*}
\begin{figure*}[h!]
    \begin{center}
    \includegraphics[width=0.99\textwidth]{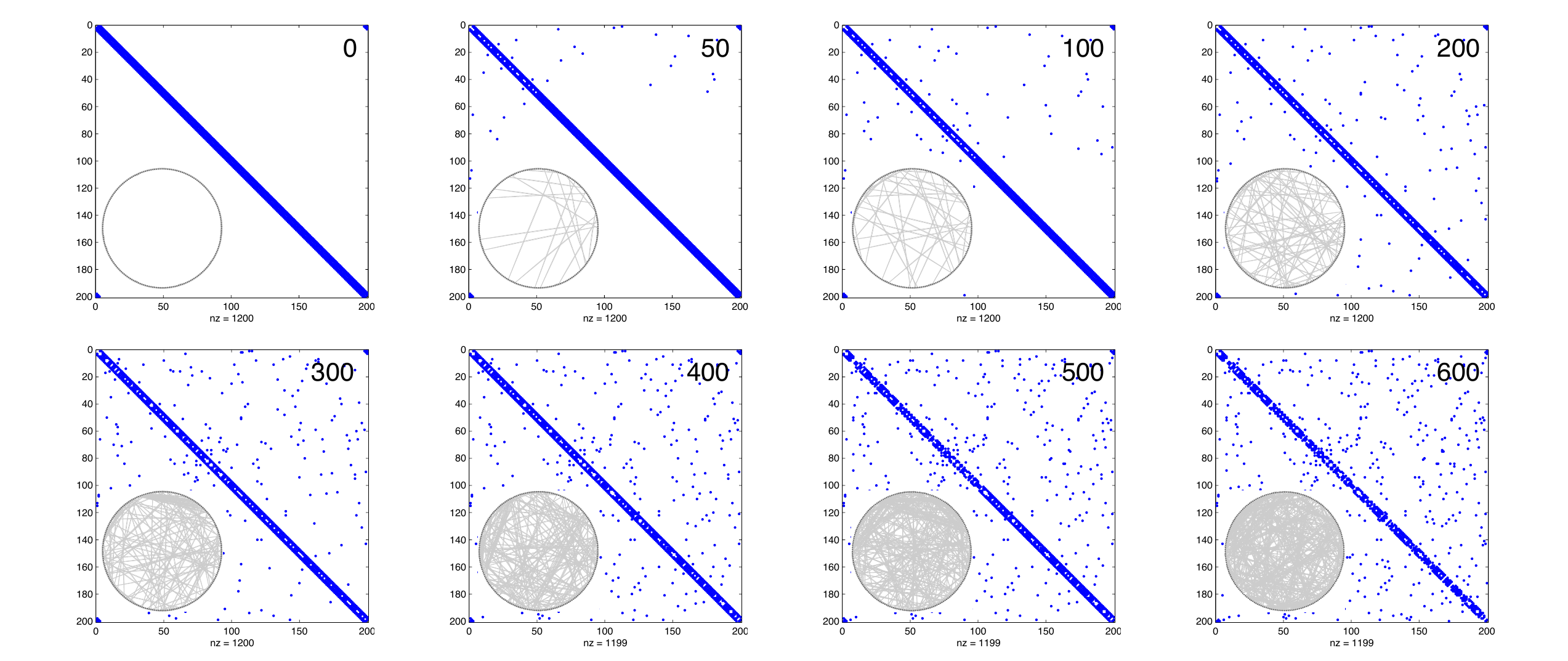}
    \caption[Re-wiring process generating a small-world network]{ \label{fig_rewiring_network}
			Adjacency matrix and belonging network during rewiring process as described by \citet{Watts1998} (number of steps in upper right corner; inset: network representation with nodes arranged on a circle).
			Beginning with a perfectly regular ring lattice (200~nodes) where each node is linked to its 6 closest neighbours (upper left), nodes are visited successively (one per step) and connections are randomly rewired with a probability of 40\%.
			On the $k$\textsuperscript{th} visit to a node, it is the link to the $k$\textsuperscript{th} neighbour on the right, which is potentially rewired.
			After 600~steps (lower right) every node has been visited three times and on average 40\% of all links have changed.
        }
    \end{center}
\end{figure*}
\begin{figure*}[!h]
    \begin{center}
    \includegraphics[width=\textwidth]{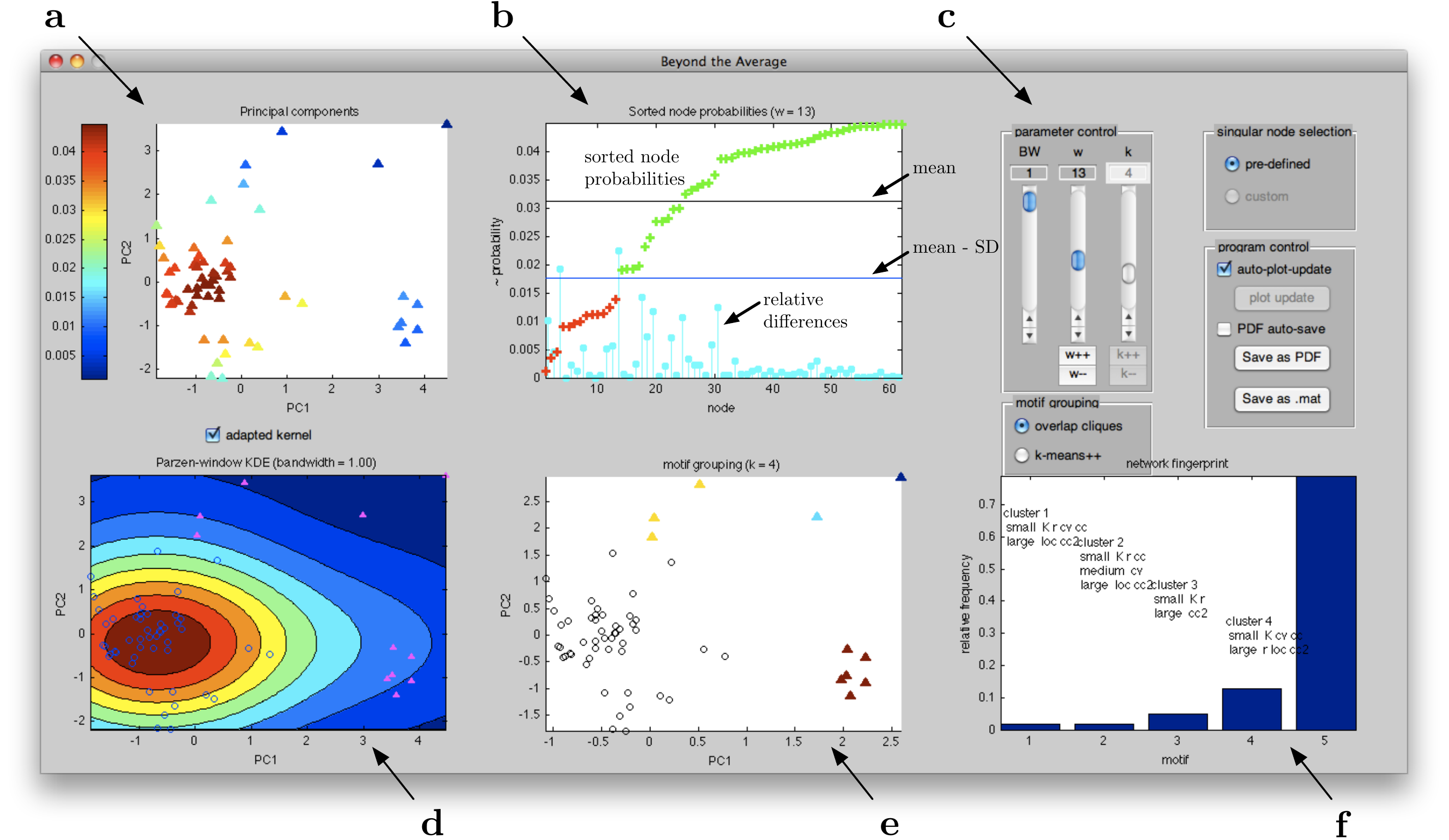}
    \caption[Graphical user interface for the BtA-workflow]{ \label{fig_GUI}
         Graphical user interface for the BtA-workflow:
         \textbf{a}~Nodes mapped to PCA-plane where their probability is coded by colour.
         \textbf{b}~Sorted node probabilities and relative differences.
         Red and green colour indicates singular and regular nodes, respectively.
         Mean probability indicated by black line; blue line marks mean minus one standard deviation.
         Stems (cyan) indicate relative differences between their two adjacent probabilities.
         \textbf{c}~Manual workflow-parameter control and options for result export.
         By default, changed settings show immediate effect in all plots (a,b,d--f).
         \textbf{d}~Contour plot of PDF with reduced feature vectors superimposed, whose colour indicates whether they are classified regular or singular.
         \textbf{e}~PCA-plane (rescaled by standard deviations) showing differently coloured motif groups.
         \textbf{f}~Bar plot showing the relative frequency for each motif-region.
         A brief characterisation of each motif is given above its bar.
        }
    \end{center}
\end{figure*}
\begin{figure}[!h]
    \begin{center}
    \includegraphics[width=0.48\textwidth]{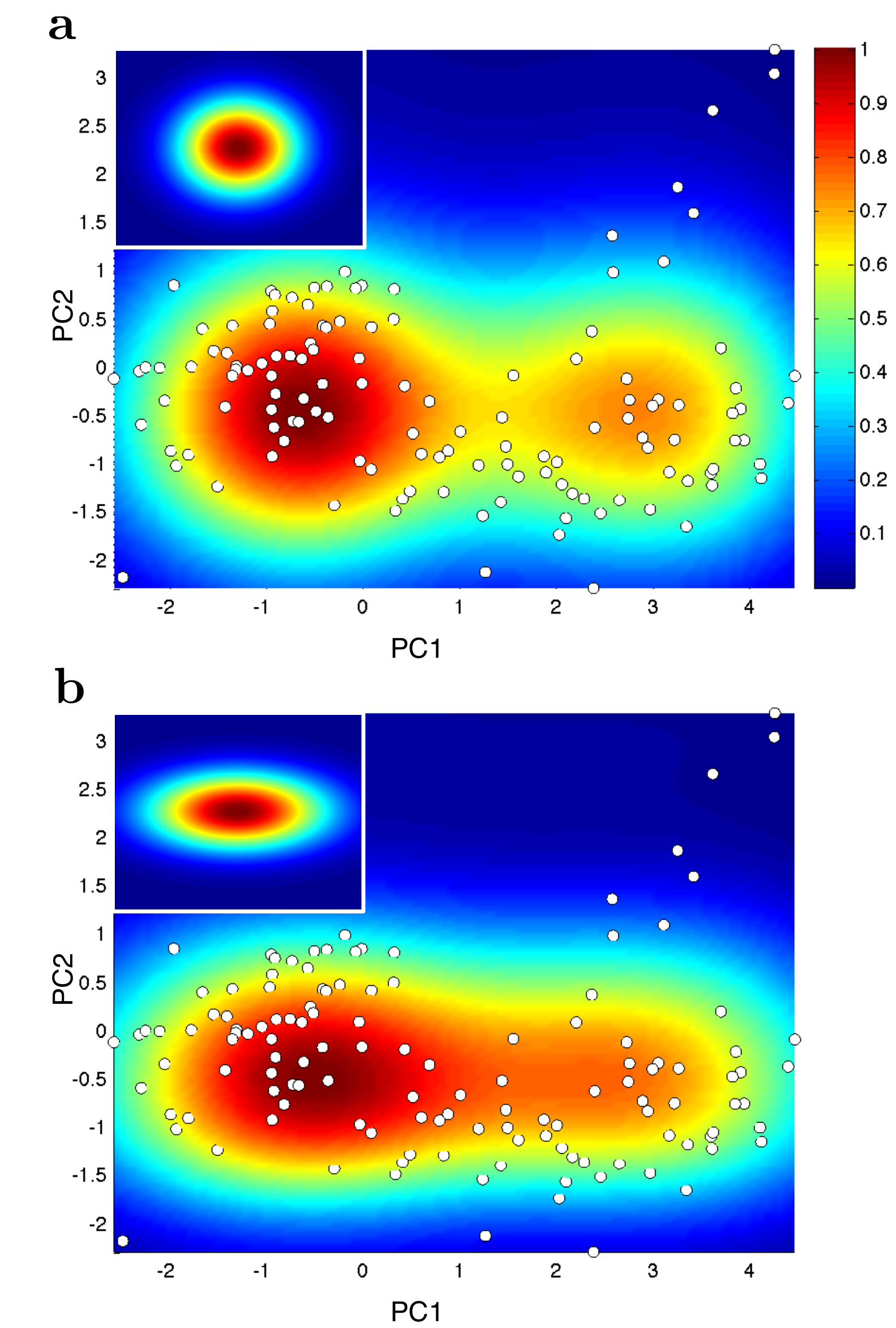}
    \caption[Contour plots of PDFs estimated using two different Gaussian kernels]{ \label{fig_adapting_kernel}
        Contour plots of PDFs estimated using two different Gaussian kernels~(upper left inlet).
        \textbf{a}~Identical kernel bandwidth in both dimensions. With this symmetric kernel, the estimated PDF shows broad spreading of probability mass along vertical axis.
        \textbf{b}~Kernel bandwidths scaled according to standard deviation along corresponding PC-axis.
        This adopted kernel results in a thinner and better matching PDF.
        }
    \end{center}
\end{figure}
\begin{figure}[!h]
    \begin{center}
    \includegraphics[width=0.45\textwidth]{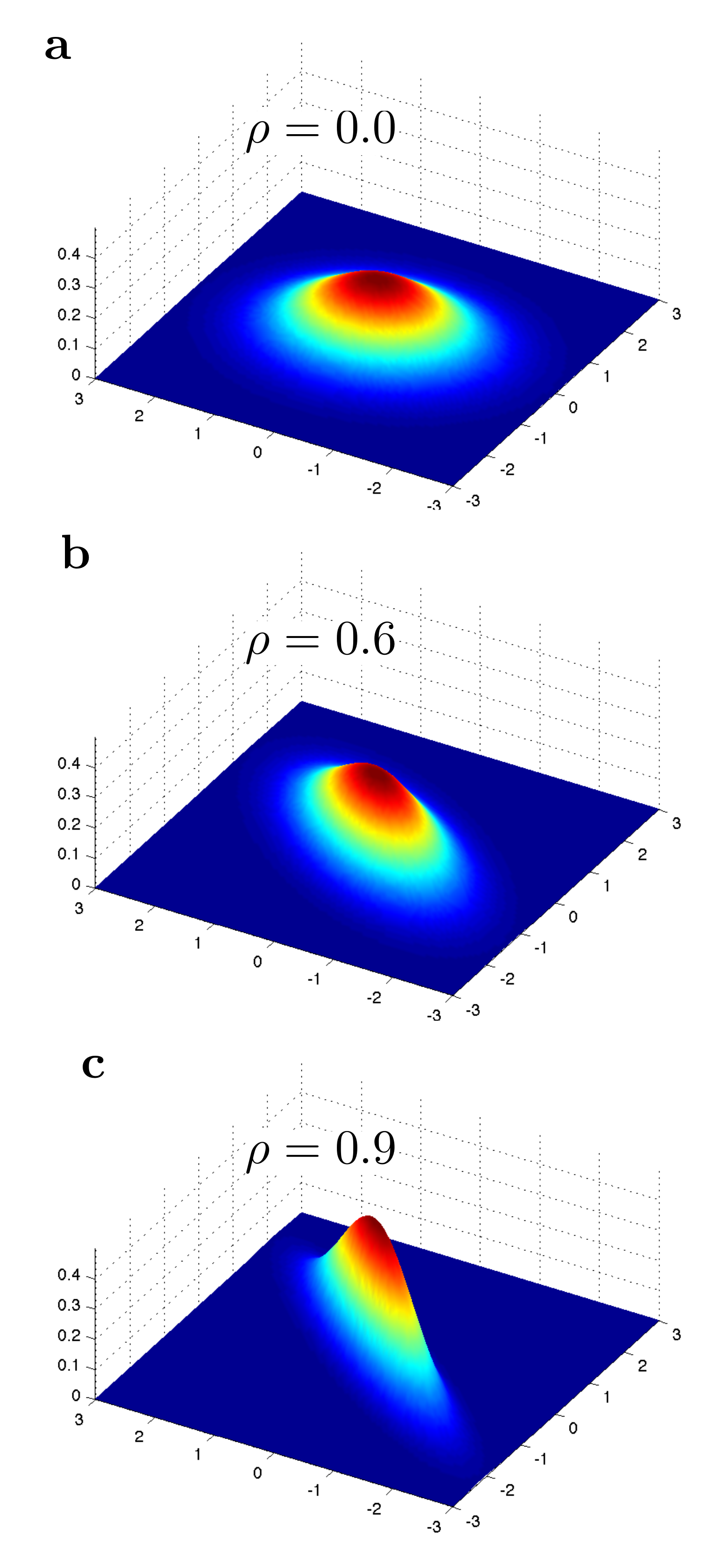}
    \caption[Density functions of 2D-Gaussian with different correlation]{ \label{fig_tilting_gaussian_3D}
         Density functions of 2D-Gaussian ($\sigma_x = \sigma_y = 1$) with different correlation ($\rho = 0.0, 0.6, 0.9$).
         \textbf{a}~Uniformity of probability mass distribution around centre ($\mu_x = \mu_y = 0$) without correlation.
         \textbf{b}~Gradually increasing correlation leads to tilting and \textbf{c}~concentration of probability mass along the diagonal.
        }
    \end{center}
\end{figure}

\clearpage
\section{Notes on the Software-Implementation}
Our implementation of the workflow including the automatic parameter determination is publicly available (\url{http://www.biological-networks.org/}).
Below we briefly mention the workflow-alternatives of the software.
\subsection{Kernel-Bandwidth} 
By default, the kernel-bandwidth is scaled according to the standard deviation along each principal component (PC) axis.
Variability-based re-shaping of the kernel function improves the overall fit of the PDF to the points (Fig.~\ref{fig_adapting_kernel}).
The kernel can also be made symmetric by deactivating the tick-box below the panel shown in Fig.~\ref{fig_GUI}a.

\subsection{Number of Singular Nodes~$w$} 
By default, our implementation of the workflow chooses the number of singular nodes~$w$ according to equation~(1).
This setting can be overwritten by the user, who is provided with a plot of all nodes' probabilities together with their relative differences (Fig.~\ref{fig_GUI}b).
Manually chosen values for~$w$ can thereby be easily related to the default setting.

\subsection{Number of Motif Groups~$k$} 
Our implementation of the workflow provides 3~alternatives to determine~$k$:
By default motif-groups are determined deterministically through cliques of overlapping ellipses, as illustrated in~Fig.~4.
The user can also choose to determine the number of clusters using the ellipses, but perform clustering with k-means++.
As the last alternative, k-means++ can be applied with a customised number of motif-groups.

\section{Run-Time Complexity}
The bulk of the runtime of the BtA-workflow is spent on step~1 where all selected local network measures are computed.
Run-time complexity here depends on the measures that are chosen to characterise each node.
We estimated how computational costs scale for six common local measures~\citep{Costa2007a}.
Like Costa et al. \citep{costa_2009}, we selected
the normalised average degree~$r$,
the coefficient of variation of the degrees of the immediate neighbours of a node~$cv$,
the clustering coefficient~$cc$~\citep{Watts1998,Kaiser2007NJP},
the locality index~$loc$,
the hierarchical clustering coefficient of level two~$cc_2$~\citep{Costa2006},
and the normalised node degree~$K$.
These measures have been applied to random networks, which have been generated according to the Erd\H{o}s-R\'enyi~(ER)~\citep{Erdos1959}, Watts-Strogatz~(WS)~\citep{Watts1998}, and Barab\'{a}si and Albert~(BA)~\citep{Barabasi1999} model.
A polynomial function was fitted (root mean square error) to the average run-times to determine their dependence on network size.
Additional to the size of the network, its edge density might also affect run-time, which is why we repeated the process while varying sparseness \footnote{%
For each random network-model (ER, BS, WS) any combination of network size ($n=10, \ldots, 100$ nodes) and edge density (10\%, 25\%, and 50\%) has been evaluated 100 times.
}.
The results show relatively stable growth rates, irrespective of network model or connection density:
Our na\"{i}ve implementations of the six measures show run-time complexities ranging from linear to less than cubic (Fig.~\ref{fig_complexity_comparison}).
Costs are thus comparatively cheap considering methods that identify specific connectivity patterns by counting occurrences of particular sub-graphs (e.g.~\citep{Wasserman1994, Kuramochi2001, Milo2002, Kashtan2004, Middendorf2005, Bordino2008}); such motif-counts also scale at least linearly in network size, but they show exponentially growing costs as the size of the motif-pattern increases~\citep{Kashtan2004}.
In practice this often means that counts can not be determined for patterns involving 10~nodes or more~\citep{Ribeiro2009}, which renders some domains computationally intractable for this approach.
In these cases the BtA-methodology might still be applicable:
Local networks measures that only scale polynomially are comparatively fast to calculate and exceptional network characteristics can therefore even be identified in very large networks.

\section{A Small-World Emerging - Detailed Result Discussion}
In total we identified 5 singular node motifs, which differ in frequency and time of emergence (Fig.~3c):
Motifs~2, 3 and~5 appear right from the beginning of the rewiring process; motifs~2 and~5 gradually become more common over time, whereas~3 levels out after a transient peak.
The remaining motifs~1 and especially~4 only become apparent at later stages towards which both become more frequent.
The motifs' temporally dependent expression levels can be understood by looking at their individual characteristics (Fig.~3d):
\begin{enumerate}

 \item A node according to motif~1 has relatively few connections in contrast to its well connected neighbourhood.
Other nodes that were initially linked to it have rewired themselves and because connections only change in 40\% of the cases, motif~1 is rarely observed in early stages.

 \item This contrasts the early appearance of motif~2 for which corresponding nodes are signified by many connections to a rather sparsely connected neighbourhood.
From the starting point of a ring lattice such configuration occurs, as re-linking one of the initial regular connections destroys the local neighbourhood structure; if multiple nodes re-wire to the same target its degree grows, which makes the node a candidate for motif~2.

 \item Motif~3-nodes have relatively few connections and nodes in their neighbourhood are similar in number of links and corresponding targets.
This characterisation fits nodes linked to others that have been disconnected from the direct neighbours only.
Such is likely to be observed during the first 200~steps of the rewiring process, where links to the closest neighbour are replaced, which is in agreement with motif~3's early peak in frequency.
Later, when connections to further away neighbours are lost, the locality index decreases and fewer nodes fulfil the profile of motif~3.

 \item The 4\textsuperscript{th} motif mostly starts to appear when nodes are visited for the third time and some of the longest initial connections are replaced.
At these late stages the ring lattice has undergone substantial perturbation, such that nodes differ widely in their degree and interconnectivity.
Motif~4 describes rarely connected nodes whose neighbours have a diverse number of connections; but instead of being linked between each other, neighbours share other common targets.

 \item The final motif~5 can be best characterised by its relation to the rest of the network, which shows a higher degree of connectivity than any node involved in the motif.
Neighbours of the motif-node further vary in their number of connections and do not link to each other.
This motif emerges early on, but its frequency rises more quickly during the last re-wiring-pass.
During that time the last initial links are broken up and motif~5 emerges, as more parts of the network finally become sparse enough.

\end{enumerate}

\clearpage
\bibliographystyle{plainnat}
\bibliography{p_citations}